# Stimulated radiative laser cooling

Peter Muys

Lambda Research Optics Europe

Tulpenstraat 2, B-9810 Eke-Nazareth, Belgium

e-mail: pmuys@lambdaeurope.be

Abstract

Building a refrigerator based on the conversion of heat into optical energy is an ongoing engineering challenge. Under well-defined conditions, spontaneous anti-Stokes fluorescence of a dopant material in a host matrix is capable of lowering the host temperature. The fluorescence is conveying away a part of the thermal energy stored in the vibrational oscillations of the host lattice. In particular, applying this principle to the cooling of (solid-state) lasers opens up many potential device applications, especially in the domain of high-power lasers. In this paper, an alternative optical cooling scheme is outlined, leading to radiative cooling of solid-state lasers. It is based on converting the thermal energy stored in the host, into optical energy by means of a stimulated nonlinear process, rather than a spontaneous process. This should lead to better cooling efficiencies and a higher potential of applying the principle for device applications.

## 1. Introduction

Numerous schemes have been proposed in recent years to realise cooling in gaseous, liquid and solid bodies through interactions of atomic particles in the body with incident optical beams. A number of methods work in atomic gases and are known under the name Doppler and sub-Doppler cooling, atomic trapping, evaporative cooling and Bose-Einstein condensation[1]. They generate temperatures in the microkelvin range or even lower. Another important category relies on spontaneous fluorescence by anti-Stokes scattering at impurity ions, to drain off excess energy and hence to lower the temperature of the ensemble of particles [2-7]. This fluorescence method works at room temperature. In this context, radiation-balanced lasers have been conceived as practical device implementation of the spontaneous fluorescence cooling principle [8]. The radiation-balancing is accomplished by offsetting the heat generated from stimulated emission by the cooling through spontaneous fluorescence. In this paper , we present a concept to cool a laser medium by a *stimulated* process , taking place in a host crystal doped with an (ionic) dopant material. Because the process consists of mixing four waves in the non-linear host matrix having a high third-order susceptibility, it will be stimulated and it can take place with much higher efficiency that the spontaneous anti-Stokes scattering process in the dopant ions which is usually invoked for fluorescence cooling. The new concept admits a broad range of specific implementations, by choosing particular combinations of pump beams, of hosts, of dopants, and of excited electronic levels of the dopant. A number of conditions need to be fulfilled to obtain macroscopic effects.

We will discuss prospects for experimental realization and practical device applications.

## 2. The dopant-host configuration

We consider a dopant element imbedded in a host matrix, typically a crystal lattice. The dopant is supposed to constitute a quasi-three level laser system, consisting of the ground level, the upper laser level and the lower laser level, which is also called the terminal laser level. All these levels refer to electronic excited states of the dopant. The terminal level is typically lying a few units of kT ( kT= 200 cm$^{-1}$ at room temperature) above the ground level of the dopant, so is thermally filled according to the Boltzmann distribution law, and corresponds to an excited state in the lower level manifold of electronic states of the dopant ion.

The upper laser level of the dopant now is directly populated by a properly tuned pump laser.

In order to implement the proposed cooling scheme, we have to fix some assumptions. The first condition is that we suppose the dopant to be imbedded in a host material disposing of a high third order non-linear susceptibility $\chi^{(3)}$. The second condition which the host has to fulfil is that it must show vibrational lattice resonances which are strongly coupled to the lower level electronic states of the dopant. This electron-phonon coupling actually occurs in classically known systems [9], e.g. between the Yb$^{3+}$-ion and YAG, where the Yb-O lattice vibration is coupled to the electronic transitions of the Yb$^{3+}$-ion, situated around 600 cm$^{-1}$ (it is implicitly assumed that the energy level of the ground state of the dopant and the host are the same). This resonant coupling is called vibronic. In this way, the terminal laser level population of the dopant is converted into phonon states of the host. Stated alternatively, the lower level excited electronic states form one common energy reservoir with the vibrational modes of the host.

## 3. The cooling principle

During regular laser operation in a quasi-three level laser, the lower laser level of the dopant is emptied by the creation (emission) of phonons in the lattice host, due to collisional de-excitation between host and dopant (non-radiative relaxation). This vibrational relaxation is described [10] by the equation:

$$\frac{dN_t}{dt} = -\frac{N_t - N_{teq}}{\tau_t} \qquad (1)$$

where $N_t$ is the lower laser level population density of the host and $N_{teq}$ represents its equilibrium value at the host temperature T. The time constant $\tau_t$ is the corresponding relaxation time. The host temperature increases as the system evolves again to thermal equilibrium, through the creation of phonons. If however, a situation can be created where the actual occupational density of the terminal level $N_t$ is *lower* than the equilibrium occupational density $N_{teq}$, at that temperature, then the system will try to re-establish thermal equilibrium in the population of the terminal level by annihilating vibronically resonant phonons from the host matrix. Hence , we have to implement a depletion channel for the terminal level population which does not pass over the ground level. To realise this, we now invoke the resonant electron-phonon coupling between dopant and host . The host phonon states are ready to participate in an anti-Stokes Raman scattering process in the host material, triggered by the pump photons and emptying in this way the vibrational level by emission of anti-Stokes photons . This scattering process however is not regular Raman scattering, where only two photons (pump and anti-Stokes) are involved. It is known that in regular stimulated Raman scattering, the anti-Stokes field will not realize amplification, whereas the Stokes field will do so [10]. On the contrary, when the anti-Stokes process makes part of a four-wave mixing process, amplification becomes possible. Two pump photons (supplied by an external laser) and one stimulated photon (supplied by the dopant) interact through the third order non-linear susceptibility $\chi^{(3)}$ of the host to create a fourth photon (see figure 1). Since the ground and phonon level of the host are involved, the scattering process in the host is resonantly enhanced. The fourth photon is the anti-Stokes photon and is emitted from the crystal.

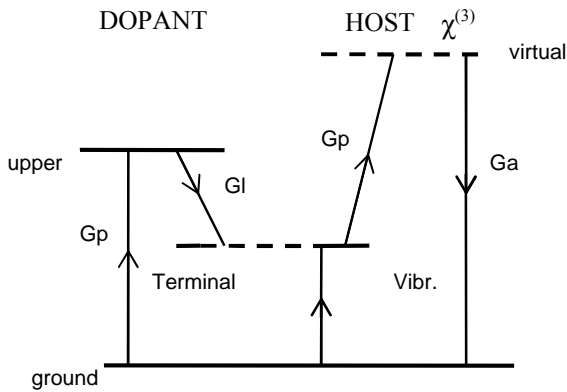

The above situation describes a strongly non-equilibrium situation for the terminal laser level. The four-wave mixing process is depleting the phonon level so efficiently that the host tries to reach thermal equilibrium again by filling the vibrational level through extraction of phonon energy from the host. This actually means that energy is conveyed away from the crystal as anti-Stokes radiation and that in this way it is being cooled radiatively.

## 4. The four-wave mixing process

For the particular four-wave mixing process considered here, the pump laser provides two photons, the dopant generates the third, and the host emits the fourth. It is clear that a part of the power of the pump beam is used to pump the dopant and another part is used to participate in the four-wave mixing process in the host. So, some pump photons are converted to stimulated photons in the dopant, and other pump photons are scattered by the host. The fourth beam annihilates phonons of the host and hence lowers its temperature, resulting in a cooling down of the system. The proposed cooling process is stimulated since it is a four-wave mixing process. This means that under the proper conditions (such as phase matching) the waves can exhibit exponential gain and hence reach much higher intensities than in spontaneous processes.

The four-wave mixing process considered here is structurally different from the regular Stokes/anti-Stokes coupling or from Coherent Anti-Stokes Raman Scattering (CARS) occuring in a non-linear host where no electronic excitations are required. In the regular CARS case, a Stokes process is occurring in conjunction with a subsequent anti-Stokes process. Both processes exclusively involve virtual states of the scattering medium. Alternatively, four-wave mixing can also take place in a medium with three excited states [10]. The four photons then correspond to four resonant transitions between these levels and the ground level. Under these circumstances, the mixing process is then called resonant. In the case considered here however, the first process is stimulated de-excitation of an excited electronic state of the *dopant*, and the second is anti-Stokes scattering in the *host*. Stated alternatively, in the first process a real excited state of the dopant is considered , whereas in the second process a virtual level of the host is involved. Host and dopant are connected here to each other by resonant phonon-electron coupling.

This four-wave mixing interaction is taking place with maximum conversion efficiency under the condition that the participating waves are phase matched in frequency and in wave vectors (conservation of energy and of momentum between the interacting photons). The first condition results in an emitted, stimulated photon with a *shorter* wavelength than the nominal laser wavelength: the anti-Stokes photon. The second condition enforces the generated anti-Stokes radiation to leave the host in the shape of a conical wave, leading to an annular power distribution in the far field. As a direct consequence, the anti-Stokes field will be diffraction-free and the exited beam corresponds to a Bessel mode beam[11]. The diffraction-free character of this field is explained by expanding it as a linear superposition of a set of plane waves all having their wave vectors lying on a cone. This operation constitutes the so-called angular spectrum expansion. It is known that in the case of regular anti-Stokes Raman Scattering , a similar situation to our case occurs [12], where also a non-collinear phase matching condition needs to be satisfied, and where the anti-Stokes beam also shows an annular shape.

## 5. Evolution of the host temperature

To analyse the temperature evolution of the ensemble, a rate equation model can be set up, describing the interaction dynamics of the different light fields with the population of the electronic levels of the dopant and with the total internal energy of the host. In the regular treatments of Raman scattering, the disturbance of the ground level by the scattering process is neglected [13]. In our model however, the interaction with the dopant levels is crucial. The host internal energy is stored in the vibrational modes of the lattice and is given by

$$E = mc_p T \qquad (2)$$

We suppose the crystal thermally isolated from its environment, so that only adiabatic changes of its internal energy can take place. The contribution of the dopant is described by the rate equations for the upper $N_u$ and the terminal $N_t$ laser level. The upper level is filled by pumping out of the ground state reservoir, and is emptied by stimulated radiation and eventually by nonradiative decay which creates phonons in the host. The terminal level is coupled to the ground level by the phonon relaxation equation (1), and is depopulated by the four-wave mixing process. In most treatises on Raman interactions, the disturbance of the ground and terminal level by the nonlinear fields are neglected [13]. In setting up a dynamical model of the system, only the photon densities are retained as variables and not the population densities of the host. In our case however, it is crucial that the interaction between all fields and all population densities are taken into account, together with the host internal energy. Photon densities for the fields are: for the pump field $\Phi_p$, for the laser field $\Phi_l$ and for the anti-Stokes field $\Phi_\alpha$. The rate equation of the laser field takes into account that light is recirculated in the cavity, by defining the cavity photon lifetime $\tau_c$. Finally, the evolution of the internal energy E is described by

$$\frac{dE}{dt} = c\ \sigma_p (N_g - N_u) G_p \Phi_p - \frac{\Phi_l}{\tau_c} + G_t \frac{N_t - N_{teq}}{\tau_t} - \Phi_a \qquad (3)$$

where $\sigma_p$ is the cross section of the pump transition, $G_p$ is the pump quantum energy and $G_t$ is the phonon quantum energy. $N_g$ and $N_u$ are the population densities of resp. the ground and upper laser level. This equation explicitly shows the heating by absorption of the pump photons, the cooling by outcoupled laser light, the nonradiative contributions by the terminal level to the phonon reservoir of the internal energy, and the draining of the generated heat carried away by the anti-Stokes beam. The temperature dependence is explicitly present in the left-hand side of eq. 3 (see eq. 1) but implicit in the right-hand side, since there $N_{teq}$ is determined by the Boltzmann distribution

$$N_{teq} = N_{dop} \frac{\exp(-G_t / kT)}{1 + \exp(-G_t / kT) + \exp(-G_u / kT)} \qquad (4)$$

$N_{dop}$ is the doping level of the host. $\Phi_a$ represents the power of the anti-Stokes beam leaving the crystal and can be called the radiative cooling term. In a regularly cooled laser, this last term is replaced by the power which is carried away by thermal conduction into the heat sink. The third contribution to eq (3) should also be negative under a far-from-equilibrium situation . Moreover, the anti-Stokes field can be written as a function of the pump and laser fields [9]. This makes the rate equation for the internal energy highly nonlinear.

In steady-state, eq. (3) equals zero, and we obtain an expression for the radiative cooling term.

In order to obtain cooling under practical working conditions, the experimental circumstances must be prepared in such a way that the probability for phonon annihilation is larger than for phonon creation. Roughly stated, this asks for hard pumping and for the availability of a highly non-linear, vibronic host. The two-level laser interaction in the dopant is also asking for hard pumping to create population inversion. So the cooling scheme will be most appropriate for pulsed, Q-switched or mode-locked lasers.

The author would like to acknowledge the advise and input of David Toebaert and Jan Muys, who teached him how to work with rate equation modelling.